# Imaging an Event Horizon:
# submm-VLBI of a Super Massive Black Hole

**A Science White Paper to the Decadal Review Committee**

**Authors:**

Sheperd Doeleman (MIT Haystack Observatory)
Eric Agol (U. Washington)
Don Backer (UC Berkeley)
Fred Baganoff (MIT)
Geoffrey C. Bower (UC Berkeley)
Avery Broderick (CITA)
Andrew Fabian (U. Cambridge)
Vincent Fish (MIT Haystack Observatory)
Charles Gammie (U. Illinois Urbana-Champaign)
Paul Ho (ASIAA)
Mareki Honma (NAOJ)
Thomas Krichbaum (MPIfR)
Avi Loeb (Harvard-Smithsonian CfA)
Dan Marrone (NRAO/U. Chicago)
Mark Reid (Harvard-Smithsonian CfA)
Alan Rogers (MIT Haystack Observatory)
Irwin Shapiro (Harvard-Smithsonian CfA)
Peter Strittmatter (U. Arizona Steward Observatory)
Remo Tilanus (JCMT)
Jonathan Weintroub (Harvard-Smithsonian CfA)
Alan Whitney (MIT Haystack Observatory)
Melvyn Wright (UC Berkeley)
Lucy Ziurys (U. Arizona Steward Observatory)
**Science Frontier Panels:**

**Galaxies across Cosmic Time**
**Cosmology and Fundamental Physics**

# A Unique Scientific Opportunity

As the most extreme self-gravitating objects predicted by Einstein's General Theory of Relativity (GR), black holes lie squarely at the intersection of Astronomy and Physics. They are now believed to reside at the heart of most galaxies, in many X-ray binary systems, and may power gamma-ray bursts: black holes are central to questions of stellar evolution, galaxy formation and mergers, astrophysical jets, and the nature of space-time. A long-standing goal in astrophysics is to observe and image the immediate environment of a black hole with resolution comparable to the event horizon. Such observations would open a new window on the study of General Relativity in the strong-field regime, astrophysics of accretion and outflow at the edge of a black hole, and fundamental black hole physics (e.g. spin).

The capabilities of Very Long Baseline Interferometry (VLBI) have improved steadily at short wavelengths, making it almost certain that *direct imaging of black holes can be achieved within the next decade.* The most compelling evidence for this is provided by a recent observation of Sagittarius A* (Sgr A*), the compact source of radio, submm, NIR and X-rays at the center of the Milky Way. SgrA* almost certainly marks the position of a ~4 million solar mass black hole, which because of its proximity has the largest apparent event horizon of any black hole candidate. This observation, using 1.3mmλ VLBI, detected structure at scales comparable to the Schwarzschild radius ($R_{sch}=2GM_{BH}/c^2$). Short wavelength VLBI of Sgr A* can and will be used to directly probe the Event Horizon of this black hole: Sgr A* is the right object, VLBI is the right technique, and this decade is the right time. Similarly, mm/submm-VLBI observations of the well-studied jet in M87 (Virgo A) will soon achieve resolutions corresponding to ~5 $R_{sch}$, enabling study of AGN jet origins with unprecedented detail.

This white paper will summarize the recent observational and technical advances that have brought event horizon imaging within reach, and then focus on four main themes and questions that can be directly addressed by high frequency VLBI observations over the next 10 years:

- Does General Relativity hold in the strong field regime?
- Is there an Event Horizon?
- Can we estimate Black Hole spin by resolving orbits near the Event Horizon?
- How do Black Holes accrete matter and create powerful jets?

## 1. Current Observations of the Galactic Center Black Hole

The case for linking SgrA* with emission from a ~$4\times10^6$ solar mass black hole at the center of the Galaxy is strong and includes: mass estimates from stars orbiting SgrA* (Gillessen et al. 2009, Ghez et al. 2008), proper motion limits on SgrA* itself (Backer & Sramek 1999, Reid & Brunthaler 2004), and size limits on SgrA* from VLBI (Bower et al. 2004, Shen et al. 2005, Doeleman et al. 2008). Taken together, results from these observations imply a mass density for SgrA* that is in excess of $9\times10^{22}$ solar masses per cubic parsec: *this is the best evidence to date for the existence of super-massive black holes (SMBH).*

High frequency VLBI allows us to see through the interstellar scattering screen that broadens radio images of SgrA* with a $\lambda^2$ dependence, while simultaneously providing angular resolutions corresponding to ~$R_{sch}$. VLBI observations of Sgr A* at wavelengths from 13cm to 3mm (Lo et al. 1985, Jauncy et al. 1989, Marcaide et al. 1992, Rogers et al. 1994, Krichbaum et al. 1998, Doeleman et al. 2001, Bower et al. 2004, Shen et al. 2005) have set important size limits on the emission region. These observations remain dominated by scattering effects, however, and the measured intrinsic sizes at these frequencies far exceed the predicted size scales associated with observed time variability of Sgr A* in the submm to xray bands (Baganoff et al. 2001, Genzel et al. 2003, Yusef-Zadeh et al. 2006, Marrone et al. 2008).

SgrA* was observed in April 2007 using a three station VLBI array (ARO/SMT, CARMA, JCMT) at a wavelength of 1.3mm (Doeleman et al. 2008). This measurement implies a size for SgrA* of only $3.7^{+1.6}_{-1.0} R_{sch}$ (3σ) assuming a circular Gaussian model (1 $R_{sch}$ = 10μas = 0.1AU). This demonstrates that we can observe SgrA* at frequencies where the intrinsic structure is not obscured by the ISM, it validates technical systems developed to perform these observations, and it confirms compact structure on event horizon scales within SgrA*. The new measured intrinsic size for Sgr A* is less than the smallest apparent size predicted for emission surrounding a non-spinning black hole (due to gravitational lensing), underscoring the ability of VLBI to probe GR in the strong-field limit (Figure 1).

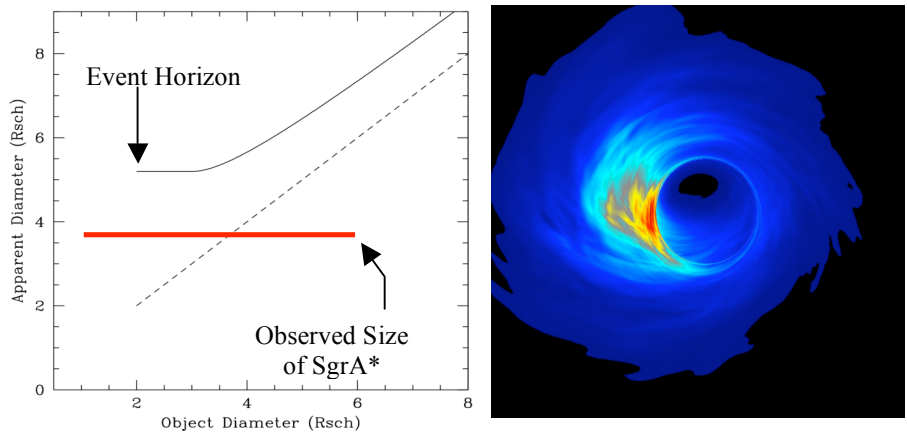

**Figure 1:** *(Left) A symmetric emitting surface surrounding a black hole is gravitationally lensed to appear larger than its true diameter. Here the apparent size is plotted as a function of the actual object size: the solid black line shows the apparent diameter with lensing by a non-spinning black hole, and the dashed line with no lensing effects included. The intrinsic size of SgrA* observed with 1.3mm VLBI (horizontal red line) is smaller than the minimum apparent size of the black hole event horizon (labeled 'Event Horizon') suggesting that the submm emission of SgrA* must be offset from the black hole position. This can be understood in the context of General Relativistic MHD accretion simulations (right), which exhibit compact regions of emission due to Doppler enhancement of the approaching side of an accretion disk (model courtesy S. Noble and C. Gammie). Jet models also produce emission peaks that are spatially offset from the black hole position (Falcke & Markoff 2000).*

These initial results are promising, but technical advances over the next decade will significantly enhance submm-VLBI observations of Sgr A*. Investments by the global astronomical community in submm facilities mean that future VLBI arrays will include many more antennas, enabling true imaging of Sgr A*. Increases in recorded bandwidth of VLBI systems will, in turn, boost array sensitivities and allow full polarization imaging. Improved VLBI frequency standards will extend VLBI to 0.8mm and 0.65mm wavelengths where interstellar scattering is negligible and array resolutions approach ~10 micro arcseconds (1 $R_{sch}$).

## 2. Using submm-VLBI to Answer Fundamental Questions of BH Physics

### 2.1. Observing Strong GR Effects: Was Einstein Right?

Einstein's theory of General Relativity has been tested and verified to high precision in the weak-field limit, including tests on Earth, observations within our solar system, and by using pulsars as extremely accurate cosmic clocks (Will 2006). Tests of GR in the strong-field regime are much more difficult since they require astronomical measurements of the most extreme cosmic environments on very fine spatial scales. Extraordinary 'movies' made with NIR adaptive optics show stars orbiting Sgr A*, with one approaching within 45AU, or some 570 $R_{sch}$, of the black hole (Ghez et al. 2008). Modeling of these orbits yields the best mass estimates for the central black hole, but these stars only sample space-time in the weak-field regime. High frequency VLBI is unique in its ability to resolve structures at the SMBH event horizon, in the case of Sgr A* and M87, and can thus be used to directly test strong-field GR predictions.

Submm-VLBI observations of SgrA* will resolve strong-field gravity effects within a few Schwarzschild radii of a black hole in the next few years. General Relativity predicts that a black hole surrounded by an orbiting or infalling optically thin plasma will exhibit a 'shadow' or 'silhouette' (Falcke, Melia & Agol 2000). This arises from two effects: emission from the near side of the BH is red-shifted and thus fainter, while emission from the far side is lensed, creating an enhanced annulus of emission near the innermost photon orbit (Figure 2 inset). The April 2007 VLBI observations can be fit with an annular structure; a non-detection on the CARMA-JCMT baseline is consistent with both the circular Gaussian and annular models (Figure 2). Planned VLBI observations, with enhanced sensitivity, will soon discriminate between these models and directly test for the 'shadow.'

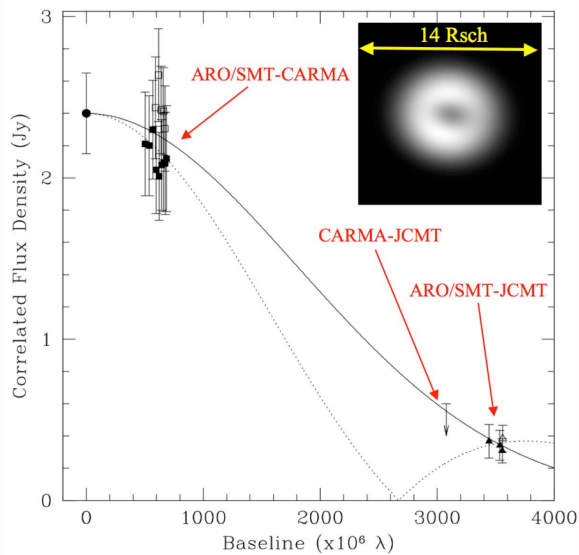

**Figure 2:** *Correlated flux density of SgrA\* as a function of baseline length from VLBI measurements in April 2007 (stations included were ARO/SMT, CARMA, and JCMT). The point at zero-baseline is a total flux density measurement made with the CARMA array. A circular Gaussian model fit with full-width-half-max of 37μas for the intrinsic size of SgrA\* is shown as a solid line. The dotted curve corresponds to what would be expected for the 'shadow' model shown in the inset, which has a 35μas inner diameter and 80μas outer diameter. Planned sensitivity increases on the CARMA-JCMT baseline will enable us to discriminate between these models – currently this baseline only provides an upper limit on the correlated flux density.*

Over the next decade the number of submm-VLBI sites will increase. Standard imaging and deconvolution techniques, as opposed to the model fitting shown above, can be employed. Figure 3 shows the progression in image fidelity as new sites are added. For the most part this effort will involve deploying VLBI instrumentation to existing sites or facilities under construction.

### 2.2. Is there an Event Horizon?

The Doeleman et al. (2008) measurement of SgrA* implies a contradiction if one assumes that there is in fact a surface within SgrA* and not an Event Horizon. On the one hand, current Radiatively Inefficient Accretion Flow (RIAF) models require mass accretion rates of $\sim 10^{-8}$ solar masses/year, and the observed bolometric luminosity of SgrA* sets a lower limit of $\sim 2 \times 10^{-10}$ solar masses/year (Yuan et al. 2003). But if SgrA* has a surface instead of an Event Horizon, then all the energy and matter not radiated away will heat this surface and it will radiate with a black-body spectrum peaking in the NIR. SgrA* is not, however, detected in quiescence in the NIR, and current NIR limits on the flux density of SgrA* set stringent upper limits on the mass accretion rate of $10^{-12}$ solar masses/year. The discrepancy between the NIR mass accretion limits and those inferred from the bolometric luminosity implies that an Event Horizon must exist in SgrA* (Broderick & Narayan 2006). New VLBI measurements at wavelengths of 1.3mm to 0.65mm will tighten this argument by resolving the structure of the emission region.

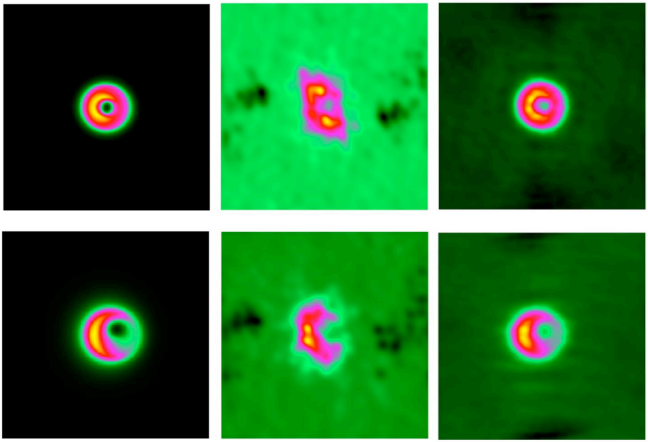

**Figure 3:** *0.8mmλ VLBI imaging simulations of Sgr A\*. The left-most images are models that have been scatter broadened by ISM effects. The middle panels show images reconstructed using a 7-station array that could reasonably be scheduled within 3-5 years. The right panels show images reconstructed using a 13-station array that could be assembled within this decade. Images on the top correspond to a GRMHD simulation with a black hole spin of a=0.5 and an accretion disk inclination of 85 degrees from our line of sight. Bottom images correspond to a RIAF model with spin a=0 and disk inclination of 60 degrees. Models courtesy Charles Gammie and Avery Broderick.*

### 2.3. Can we estimate Black Hole spin by resolving orbits near the Event Horizon?

Sgr A* exhibits variability in the x-ray and NIR on time scales that correspond to a light-crossing distance of few $R_{sch}$, and Sgr A* is seen to flare in the submm about once every 1-2 days (Yusef-Zadeh 2006, Eckart et al. 2006, Marrone et al. 2008). It is expected that during these ~1-2 hour submm flares the small scale structure of Sgr A* will change, with the implication that long duration VLBI observations will be unable to image the emission without significant smearing of the resulting image. One model for such activity invokes localized heating in the surrounding accretion flow, potentially due to magnetic reconnection, resulting in 'hot-spots' (Broderick & Loeb 2006). Estimates suggest that these 'hot-spots' could persist for multiple orbits. The period of the Innermost Stable Circular Orbit (ISCO) for Sgr A* ranges from ~30 minutes for a zero spin black hole to ~4 minutes for maximal spin (Frolov & Novikov 1998); within the ISCO matter quickly plunges inward to the event horizon. Genzel et al. (2003) claim to have observed a ~17 minute periodicity in NIR flux during a SgrA* flare, which would imply a spin of half the maximal rate, but a recent analysis of a longer light curve (Meyer et al. 2008) finds no significant periodicity. *Submm-VLBI over the next few years should be able to measure the ISCO period by resolving matter as it orbits the black hole.*

Though real-time imaging of such structures is not possible, VLBI can be used to monitor and model time variable emission on $R_{sch}$ scales through non-imaging analysis of 'closure quantities'. These are sums and ratios of interferometric phases and amplitudes computed over closed loops of VLBI baselines that are robust against calibration errors (Rogers, Doeleman & Moran 1995). Figure 4 shows that VLBI closure phase information can follow complex structures, in this case those predicted by a model hot-spot in orbit around SgrA*. By extracting the periodicity of these orbits, the black hole spin can be estimated (Doeleman et al. 2009). The effect of including the ALMA array in these observations can be seen in Figure 5, which shows simulated 10-second closure phases that include data on long VLBI baselines from the US to Chile. Increasing the collecting area at the Chilean site from a single dish to a phased sum of 10 (or more) ALMA dishes allows detailed modeling of the orbital structure as well as robust extraction of the orbital period. The exact nature of the flaring structure will undoubtedly be more complex than these simple models, but submm-VLBI can effectively probe size and time scales on which flares occur.

The fine time resolution of this promising technique draws comparison with future plans to monitor Iron fluorescence lines from 'hot-spots' using the International Xray Observatory satellite (IXO) (Brenneman et al. 2009). Submm-VLBI observations of flares in Sgr A* will take place throughout this decade, thus providing unique and valuable information for the design and planning of future space missions.

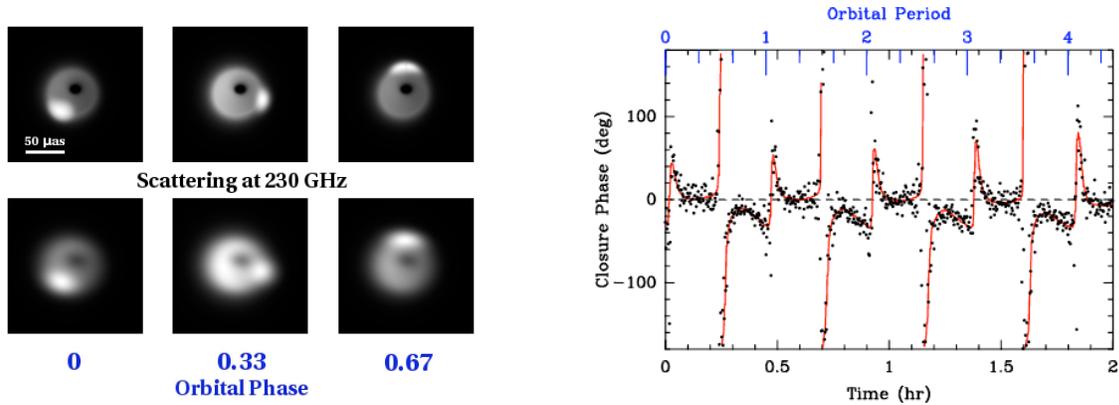

**Figure 4**: *Signature of a hot-spot orbiting the SgrA\* black hole. The left panel shows a quiescent Radiatively Inefficient Accretion Flow (RIAF) model for a non-spinning $4\times10^6$ solar mass black hole, and a hot spot orbiting at the Innermost Stable Circular Orbit (ISCO), with a disk inclination of 60 degrees from line of sight. The raw model is shown for 3 orbital phases in the top three figures, and the bottom three show the effects of scattering by the ISM. VLBI closure phase (the sum of interferometer phase over a triangle of baselines) is non-zero when asymmetric structure is present. The right panel shows 1.3mm wavelength VLBI closure phases every 10-seconds on the ARO/SMT-Hawaii-CARMA triangle with the model phases shown as a red curve (Doeleman et al. 2009).*

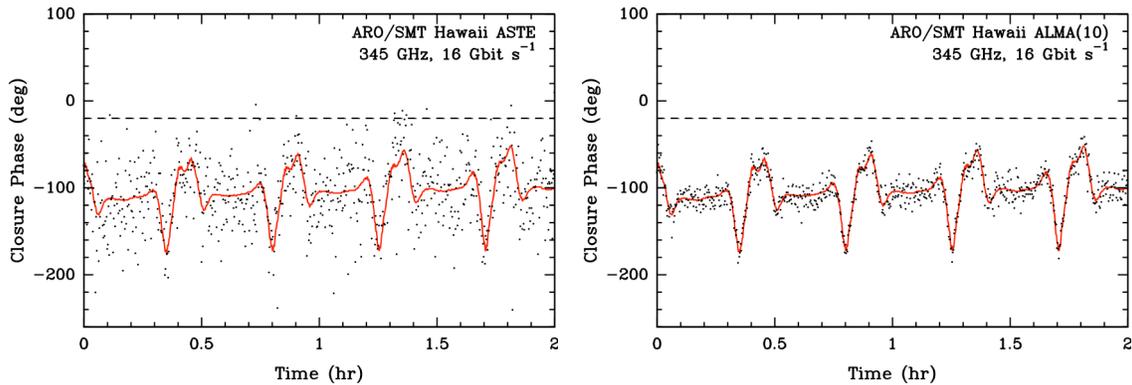

**Figure 5:** *Signature of a hot-spot orbiting an a=0.9 spin black hole at a radius of 3Rsch with a period of 27 minutes. The left panel shows the expected closure phase from the model in red and 10 second closure phase points on a 0.8mm wavelength VLBI array consisting of ARO/SMT, Hawaii (CSO+JCMT+SMA), and the ASTE dish in Chile (thermal noise has been added in these simulations. Although a non-zero average closure phase can be discerned, the points cannot be matched closely to the model. The right panel shows the effect of replacing the single ASTE aperture with 10 phased ALMA dishes (possible within 2-4 years). Now the 10 second closure phase points follow the expected model very closely, allowing the clear periodicity to be reliably extracted and the black hole spin to be estimated (Doeleman et al. 2009).*

### 2.4. How do Black Holes accrete matter?

Models of the pan-chromatic emission from SgrA* must consistently account for the black hole mass and spin, radiation mechanisms, and radiative transfer along geodesics. Increasingly sophisticated simulations make use of new algorithms (e.g. MHD turbulence: Noble et al. 2009, GR ray-tracing algorithms: Huang et al. 2008) to meld the above effects into time variable synthetic images, which can be used to predict VLBI results. Radiatively Inefficient Accretion Flow (RIAF) models explain the low luminosity of SgrA* ($\sim 10^{-8}$ of Eddington) by convecting some accreted matter away in outflows while some matter and energy pass through the Event Horizon (Yuan et al. 2003). General Relativistic MHD simulations attempt to model the spectrum and structure of SgrA* by evolving the magnetic fields and plasma surrounding the black hole (Noble et al. 2006). Jet models of SgrA* use scaled versions of the more powerful jets observed in bright extragalactic AGN (Falcke & Markoff 2000). For all of these models, the simulations and theory have hitherto far outpaced the observations, but over the next decade observations using VLBI arrays with increased sensitivity and resolution will dramatically constrain the models.

Submm-VLBI of Sgr A* represents a singular opportunity to test black hole accretion and outflow modes on the finest scales. Broderick et al. (2009), for example, have fit a generalized RIAF model to the 1.3mm VLBI data, in which the emission is parameterized by black hole spin, accretion disk inclination, and orientation of the source on the sky. They find that RIAF disk inclinations within ~30 degrees of 'face-on' can be excluded since they predict sizes in excess of that observed. Future submm-VLBI observations will improve parameter estimation in such models by incorporating additional VLBI sites whose positions on the Earth let us sample source structure more completely (Fish et al. 2009). More broadly, future VLBI observations will guide and shape next-generation computational and phenomenological models of black hole accretion by providing $R_{sch}$-scale observations of polarization structures (through full polarimetric imaging) and time-variable morphology.

### 2.5. How do Black Holes create powerful jets: M87?

Large-scale relativistic jets are a spectacular manifestation of the compact engines that power extragalactic AGN. On kiloparsec scales, the structure of these jets depends sensitively on interaction of the relativistic particle stream with the galactic ISM, with some jets such as M87 (Virgo A), exhibiting severe bending before culminating in large radio lobes (Biretta, Owen & Hardee 1983). Close to the galactic core the inner jet is presumably determined less by environment and more by the structure of the central engine. Submm-VLBI offers a new way to study the launching region of these jets with unprecedented angular resolution.

HST observations show that M87's black hole has $M_{BH} \sim 3 \times 10^9$ Msol (Macchetto et al. 1997), so Rsch = $9 \times 10^{14}$ cm ~ 60 AU. At a distance of 16 Mpc (Tonry et al. 2001), this size subtends an angle of 3.7µas, making M87 the second largest black hole candidate in terms of apparent size. Models of jet formation typically invoke magnetic fields that are either frozen into a rotating accretion disk or anchored in the ergosphere of a spinning black hole. Both mechanisms concentrate and twist the magnetic fields into helical coils that accelerate and collimate plasma away from the central black hole (e.g. Livio 1999). Thus the size of the inner edge of the accretion disk – where the magnetic fields are expected to be strongest – and the black hole ergosphere are fundamental length scales. These scales are set by the ISCO, which will be gravitationally lensed to an apparent diameter 7.4 Rsch, or ~27µas, for the zero-spin case, and 4.5 Rsch, or ~17µas, for a maximally spinning black hole.

Submm-VLBI of the M87 core will set a limit on the size of the jet base, and provide new constraints for numerical jet models. Nishikawa et al. (2005), for example, produce GRMHD jet models using an inner disk radius of 3 Rsch, corresponding to the non-spinning ISCO. De Villiers et al. (2003), in contrast, find consistent models using an inner radius of 15 Rsch in the context of a spinning black hole. Such differences are anything but incidental. Physical conditions within the inner accretion flow and jet-launching region such as the Alfvén speed, sound speed, and magnetic field depend critically on the assumed size of the inner disk. Simply detecting M87 at 1.3mm wavelength on the Hawaii-Chile baseline, would set a size limit of <17μas, tightly constraining the size of the jet formation region for M87. Broderick & Loeb (2009) have shown that in the case of M87, the black hole 'shadow' can potentially be observed and imaged.

## 3. Summary

Over the next decade, existing and planned mm/submm facilities can be combined into a high sensitivity, high angular resolution "Event Horizon Telescope", capable of imaging a black hole. This effort will include development and deployment of submm dual-polarization receivers, highly stable frequency standards to enable VLBI at 1.3mmλ to 0.65mmλ, higher bandwidth VLBI backends and recorders, and commissioning of new submm VLBI sites. Development of phased array technology to combine all submm apertures on Mauna Kea into a single VLBI station is already underway and will be adapted for use at ALMA and CARMA. We emphasize that the path forward is clear, and recent successful observations have removed much of the risk that would normally be associated with such an ambitious project. Details of the technical efforts required to assemble this "Event Horizon Telescope" will be described elsewhere, but no insurmountable challenges are foreseen.